\documentclass[%
aip,jap,
 amsmath,amssymb,floatfix,
 preprint,%
]{revtex4-1}
\usepackage{amsmath,amssymb,amsfonts}
\usepackage{algorithmicx}
\usepackage{mathtools}
\usepackage{algorithm}
\usepackage{algpseudocode}
\usepackage{graphicx}
\usepackage{textcomp}
\usepackage{xcolor}
\usepackage{mathtools}
\DeclarePairedDelimiter\abs{\lvert}{\rvert}%

\def\BibTeX{{\rm B\kern-.05em{\sc i\kern-.025em b}\kern-.08em
    T\kern-.1667em\lower.7ex\hbox{E}\kern-.125emX}}
\begin{document}

\title{$n$-bit Data Parallel Spin Wave Logic Gate\\
{}
\thanks{ This work has received funding from the European Union's Horizon 2020 research and innovation program within the FET-OPEN project CHIRON under the grant agreement No. 801055.}
}

\author{Abdulqader Mahmoud}
\affiliation{Delft University of Technology, Department of Quantum and Computer Engineering, 2628 CD Delft, The Netherlands}

\author{Frederic Vanderveken}
\affiliation{KU Leuven, Department of Materials, SIEM, 3001 Leuven, Belgium}
\affiliation{Imec, 3001 Leuven, Belgium}

\author{Florin Ciubotaru}
\affiliation{Imec, 3001 Leuven, Belgium}

\author{Christoph Adelmann}
\affiliation{Imec, 3001 Leuven, Belgium}

\author{Sorin Cotofana}
\affiliation{Delft University of Technology, Department of Quantum and Computer Engineering, 2628 CD Delft, The Netherlands}

\author{Said Hamdioui}
\affiliation{Delft University of Technology, Department of Quantum and Computer Engineering, 2628 CD Delft, The Netherlands}

\begin{abstract}
Due to their very nature, Spin Waves (SWs) created in the same waveguide, but with different frequencies, can coexist while selectively interacting with their own species only. The absence of inter-frequency interferences isolates input data sets encoded in SWs with different frequencies and creates the premises for simultaneous data parallel SW based processing without hardware replication or delay overhead. In this paper we leverage this SW property  by introducing a novel computation paradigm, which allows for the parallel processing of $n$-bit input data vectors on the same basic SW based logic gate. Subsequently, to demonstrate the proposed concept, we present $8$-bit parallel $3$-input Majority gate implementation and validate it by means of Object Oriented MicroMagnetic Framework (OOMMF) simulations. To evaluate the potential benefit of our proposal we compare the $8$-bit data parallel gate with equivalent scalar SW gate based implementation. Our evaluation indicates that $8$-bit data $3$-input Majority gate implementation requires $4.16$x less area than the scalar SW gate based equivalent counterpart while preserving the same delay and energy consumption figures. 
\end{abstract}

\maketitle

\section{Introduction}
The information technology revolution our society experienced during the past decade resulted in the generation of huge amounts of raw data, which precessing requires efficient computing platforms \cite{data2}. So far, CMOS downscaling was able to meet these requirements, however, due to different technological issues: (i) reliability wall \cite{cmosscaling1}, (ii) leakage wall \cite{cmosscaling2,cmosscaling3}, and (iii) cost wall  \cite{cmosscaling1,cmosscaling2}, CMOS downscaling turns out to be more in more difficult, which indicates the close end of Moore's law. In view of this, different technologies, e.g., tunneling FET, Graphene pn-Junction, spintronics, and memristor, have been considered as CMOS alternatives to meet the data processing market requirements \cite{survey2}. 

Among those Spin-Wave (SW) based computing seems be one of the most promising avenues due to its \cite{survey2}: (i) Ultra-low power consumption (it doesn't involve any charge movements),  (ii) acceptable delay, (iii) scalability to nm wavelength range, and (iv) intrinsic data parallelism (multiple frequencies can independently and simultaneously coexist in the same waveguide).

In view of SW great potential, various logic gates based on SW interactions have been reported, e.g., \cite{logic21,logic12,logic11,logic17,Magnonic_transistor,logic24, logic1, logic19,logic100,logic101,parallel_data_processing1}. The first SW logic gate  experimentally realized was a current controlled Macha-Zender interferometer based  NOT gate  \cite{logic21}. Subsequently, by using a similar approach, XNOR, NAND and NOR logic gates were designed \cite{logic12,logic11,logic17}. Two parallel re-configurable nano-channel magnonic devices were used to design voltage-controlled XNOR and NAND gates \cite{logic24} and by placing two magnon transistors in the arms of a Mach-Zehnder interferometer, an XOR gate was designed \cite{Magnonic_transistor}. While the previous gates make use of SW amplitude information encoding SW phase information encoding has been also considered \cite{logic1} and different logic gates including buffer, inverter, (N)AND, (N)OR, XOR and Majority gates were proposed  \cite{logic1}. Moreover, cross structures were used to build (N)OR gate \cite{logic19}. Additionally, two phase encoding based Majority gates were physically realized  \cite{logic100} \cite{logic101}.  

While the previously mentioned proposals disregard (iv), which potentially opens a data parallel computation avenue unfeasible for CMOS and in general for any charge moving based technology,  in has been suggested in \cite{parallel_data_processing1} that one $3$-input Majority gate can simultaneously process $3$ input data sets encoded in SW with different frequencies. 
The presented structure contains bent regions, which are not preferred in SW designs, and makes use of magnonic crystals as input and output filters, which induces substantial delay overhead.

In this paper we revisit the multi-frequency SWs  support for data parallelism, discuss it in its general form, and propose a generic $n$-frequency data parallel in-line gate structure. Subsequently, we present XOR and Majority SW gates capable to simultaneously process  in the same waveguide $8$ input data sets encoded in $8$ different frequencies. The main contributions of this work can be summarized as follows:\\
\begin{itemize}
\item Design of multi-frequency byte-wide in-line Spin Wave logic gate: $8$-bit $3$-input Majority gate is implemented using the multi-frequency in-line structure.
\item Validation of the proposed structure: The byte-wide Majority gate is validated by means of OOMMF simulations.
\item Comparison of the design with the conventional approach: The proposed byte-wide 3-input Majority gate requires 4.16x less area than the conventional approach with similar latency and energy consumption.
\end{itemize}

The rest of the paper is organized as follows. Section \ref{sec:Basics and background of SW technology} provides basic  background of SWs physics and SW based computing. Section \ref{sec:Proposed Parallelism Structure} describes the proposed data parallel gate and Section \ref{sec:Simulation setup and experiments} provides inside on the simulation setup, and parameters. In Section \ref{sec:Simulation results and discussion} we evaluate $8$-bit instance of the proposed design, compare it with conventional scalar logic gate based counterpart, and discuss scalability aspects. Section \ref{sec:Conclusion} concludes the paper.

\section{SW Based Computing Background}
\label{sec:Basics and background of SW technology}

The electron spins tend to align themselves along the applied magnetic field direction to decrease the total energy to the lowest level, when applying an external magnetic field to a ferromagnetic material \cite{Magnonic_crystals_for_data_processing}. If an, e.g., Magnetoelectric (ME) cell, antenna, based excitation method is utilized to deflect the electron spin, a Spin Wave (SW) is created  by exchange and dipole interactions. This results in a precessional spin movement \cite{Magnonic_crystals_for_data_processing}, which can be described by  the Landau-Lifshitz-Gilbert (LLG) equation \cite{LL_eq}:
\begin{equation} \label{eq:1}
\frac{d\vec{m}}{dt} =-\abs{\gamma} \mu_0 \left (\vec{m} \times \vec{H}_{eff} \right ) + \alpha \left (\vec{m} \times \frac{d\vec{m}}{dt}\right ),
\end{equation}
where $\alpha$ is the damping factor, $\gamma$ the gyromagnetic ratio, $\mu_0$ the vacuum permeability, and $m$ the magnetization. Also, $H_{eff}$ is the effective field and it is equal to $H_{eff}=H_{ext}+H_{ex}+H_{demag}+H_{ani}$ where $H_{ext}$ is the external field, $H_{ex}$ the exchange field, $H_{demag}$ the demagnetizing field, and $H_{ani}$ the magneto-crystalline anisotropy.

An excited SW is characterized by its $f$, determined by the time a complete spin precession takes,  wavelength $\lambda$, the shortest distance between two spins exhibiting the same spinning behaviour, wave-number $k$ $\left(k=\frac{2*\pi}{\lambda}\right)$, amplitude $A$, and phase $\phi$. Such SWs can simultaneously propagate through the same waveguide, while carrying information potentially encoded in amplitude, phase, frequency or combination of those, and their mutual interaction is governed by the superposition and interference principles. As such, two SWs having the same $A$, $\lambda$, and $f$  will constructively or destructively interfere if their phase difference is $0$ or $\pi$, respectively. Moreover, the majority decision governs the interference process when more than two waves with the same $A$, $f$, and $\lambda$ coexist in the waveguide. Specifically, if the number of SWs that have $\phi = 0$ (logic "0") is greater than the number of SWs that have $\phi = \pi$ (logic "1"), then the interference result is logic $0$ and $1$ otherwise, which provides a natural support for the direct evaluation of Majority functions. Consequently, the behaviour of a $3$-input Majority gate, which CMOS implementation requires $18$ transistors, can be mimicked by the interference in the same waveguide of $3$ SWs  \cite{logic1}.  In addition, SWs carrying amplitude and/or phase encoded information but with different $f$s can simultaneously propagate in the same waveguide while only interfering with their own species (waves having the same frequency), which provides natural supports for data parallel computing. We note that, in the most general case, SW with distinct $A$, $f$, $\phi$, and $\lambda$ generated in the same waveguide will interfere and generate SWs results that can provide support for yet to be unveil computation paradigms, but in this paper we only concentrate on the interaction of SW with the same amplitude and frequency. 

Different SW types, i.e., exchange spin wave, exchange-dipole spin wave, magnetostatic
surface spin wave, forward volume magnetostatic spin-waves, backward volume magnetostatic spin-waves, can be excited depending on the SW propagation direction versus the magnetization and effective magnetic field directions \cite{Magnonic_crystals_for_data_processing}. While each type has certain attractive characteristics, as seen from the circuit design point of view Forward Volume Magnetostatic spin-waves (FVMSWs) appear to be the most interesting  because the in-plane spin-wave propagation is isotropic. As a result, the same wave-number is generated in all directions, which creates circuit design opportunities not achievable for the other SW types. 
\section{$n$-bit Data Parallel SW Logic Gate}
\label{sec:Proposed Parallelism Structure}

\begin{figure}[t]
\centering
  \includegraphics[width=0.85\linewidth]{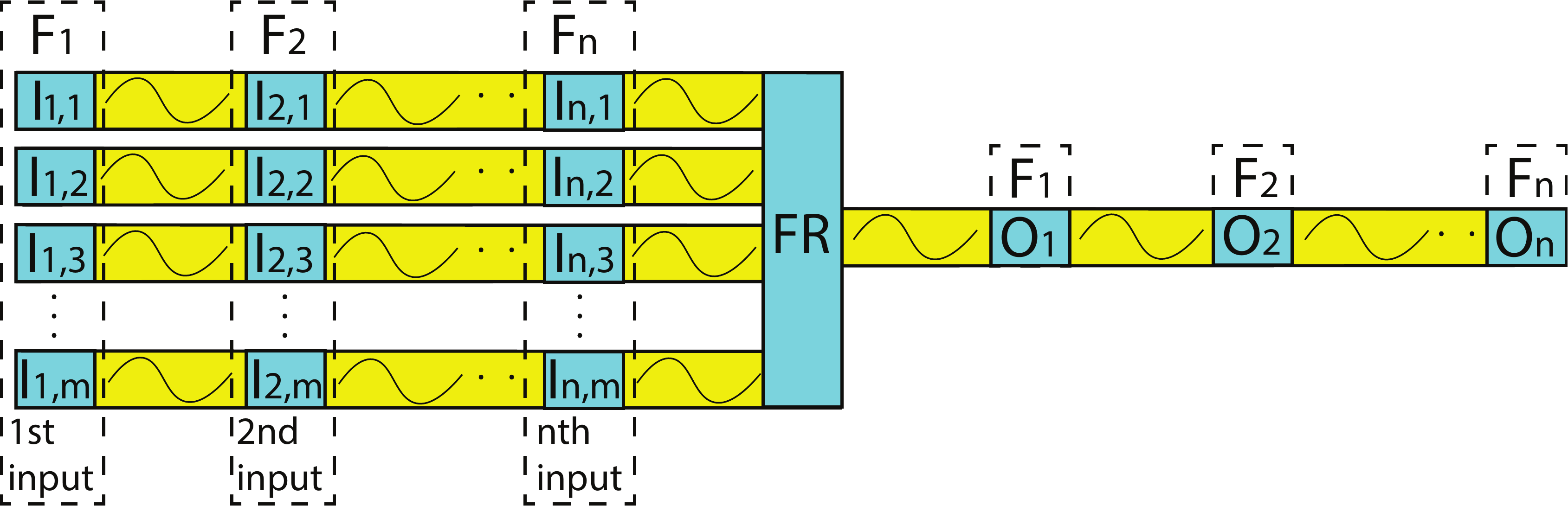}
  \caption{Multi-Frequency Spin Wave Logic Gate}
  \label{fig:proposed_architicture}
\end{figure}

\begin{figure}[b]
\centering
  \includegraphics[width=\linewidth]{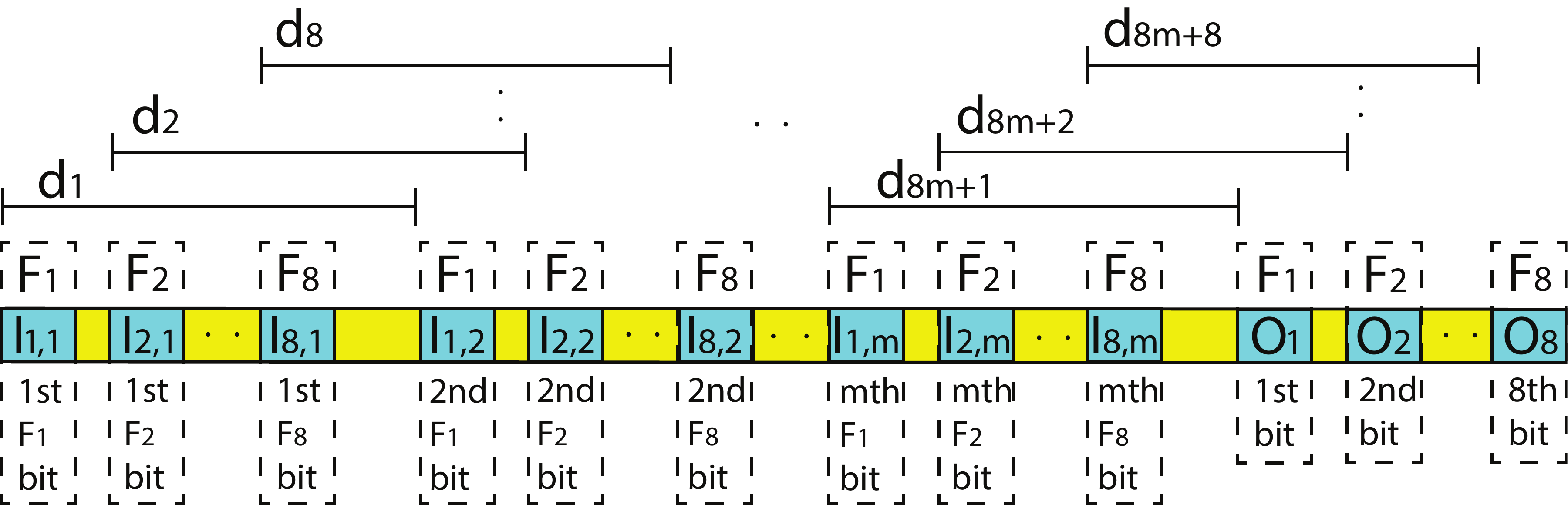}
  \caption{Byte-based In-line Spin Wave Logic Gate Structure}
  \label{fig:proposed_structure}
\end{figure}

The proposed $m$-input $n$-bit data parallel SW logic gate structure is depicted in Figure \ref{fig:proposed_architicture}. It operates on a set of $n$ $m$-bit values, while each set element is being encoded in it own frequency, $f_i, \{i=1, 2, \ldots, n\}$. As graphically indicated in the Figure, SW excitation elements, e.g., ME cells, placed along the $m$ input waveguides are utilized to transform the input values at  $I_{1,1}$, $I_{1,2}$, $I_{1,3}$, \ldots, $I_{m,n}$ into SWs, which subsequently propagate towards the Functional Region (FR) without interfering with each-other. Only when the $m \times n$ waves reach the FR SWs within each group \{$I_{1,1}, I_{1,2}, I_{1,3}, \ldots, I_{m,i}\}_{i=1,n}$ start to interact, which results in $n$ output SWs that can be read by transducers placed at $O_1$, $O_2$, \ldots, $O_n$ or passed to potential following SW gates. As the proposed structure enables independent propagation and interaction of multiple frequency SWs it allows for parallel processing of multiple input sets within the same structure. This, potentially results in delay and area savings when compared to a conventional evaluation based on serialization or hardware replication, respectively.

While the structure is generic, its implementation by following the topology depicted in Figure \ref{fig:proposed_architicture} relies on waveguide stacking on top of each other and have bent regions, which are impeding proper SW propagation. To this end, a more appropriate realization is the one waveguide only in-line structure depicted in Figure \ref{fig:proposed_structure} for the particular case of an $8$-input $8$-bit data parallel gate. 
As indicated in the Figure the voltage- or current-encoded logic values ($I_{1,i}$, $I_{2,i}$, \ldots, $I_{8,i}$) applied to ME cells or other transducers are used  to excite  $f_i, \{i=1, 2, \ldots, 8\}$ SWs within the same waveguide, which eventually interact and generate an $8$-bit output. 

To ensure correct gate functionality, SWs with the same frequency must be excited with the same amplitude and wavelength. In addition, the distance between the SW excitation location and the interference points have to be properly adjusted as they play a key role in the gate behaviour. For example, if the desired result of SWs interference is constructive if they have the same phase and destructive if they are out of phase, then the distances between similar frequency sources must be $n \times \lambda_i, i=(1,2,3, \ldots, 8)$, i.e, $d_1 = n\lambda_1$, $d_2=n \lambda_2$, \ldots $d_8=n \lambda_8$ (where $n=1, 2, 3, \ldots$). In contrast, if the opposite behaviour is required, the distances between similar frequency sources must be $\frac{n}{2} \lambda_i$, i.e., $d_1=\frac{n}{2} \lambda_1$, $d_2=\frac{n}{2} \lambda_2$, \ldots, $d_8=\frac{n}{2} \lambda_8$  (where $n=1, 3, 5, \ldots$).

Due to the very SW nature the gate can provide direct or complemented output values by properly adjusting the position, i.e., the transducer location, at which each output is read.  For example, if the non-inverted output is required the transducers sould be located at  $n \lambda_i, i=(1,2,3, \ldots, 8), n=(1,2,3, \ldots)$ from the last $f_i$ SW source, i.e, $d_{n+9}=n \lambda_1$, $d_{n+10}=n \lambda_2$, ... $d_{n+16}=n \lambda_8$. When the inverted version is of interest the output should be detected at a distance  $\frac{n}{2}$ $\lambda_i$  from the last $f_i$ SW source, i.e., $d_{n+9}=\frac{n}{2} \lambda_1$, $d_{n+10}=\frac{n}{2} \lambda_2$, ... $d_{n+16}=\frac{n}{2} \lambda_8$ (where $n=1,3,5, \ldots$). The two strategies can be combined in case that the direct function is required for some input sets and the inverted for the rest.

\section{Experimental Setup}
\label{sec:Simulation setup and experiments}
This section provides inside on the simulation platform and parameters, we utilized to validate our proposal. 

\subsection{Simulation Platform}
To validate the proposed structure and get inside on its functionality, the Object Oriented MicroMagnetic Framework (OOMMF) \cite{OOMMF}, which predicts the magnetization dynamics by numerically solving the LLG equation, was utilized. OOMMF inputs consists of Tckl/Tk scripts describing the gate structure and for a better visualization, we Matlab post-processed  the OOMMF simulation results. %\medskip\\
\subsection{Simulation Parameters}
%OOMMF \cite{OOMMF} was used to test and validate the proposed structure. 
During the experiments we made use of a $Fe_{60}Co_{20}B_{20}$ waveguide with Perpendicular Magnetic Anisotropy (PMA), with has an anisotropy field $H_{anisotropy} > M_s$, which implies that no external magnetic field is required. The waveguide thickness and width were set to $1$ nm and $50$ nm, respectively. To evaluate the gate structures, the following parameters values were utilized during the simulations: perpendicular anisotropy constant $k_{ani}$ = $8.3177$ $\times$ $10^5$ J/$m^3$, magnetic saturation $M_s$ = $1.1$ $\times$ $10^6$ A/m, damping constant $\alpha$ = $0.004$, and exchange stiffness $A_{exch}$ = $18.5$ pJ/m \cite{parameters}. As we evaluated $8$-bit data parallel gates we made use of SWs with $10$ GHz, $20$ GHz, $30$ GHz, $40$ GHz, $50$ GHz, $60$ GHz, $70$ GHz, and $80$ GHz frequencies. Based on the FVMSW dispersion relation and the wavenumber $k=\frac{2\pi}{\lambda}$, the distances between the same frequency sources were determined as follows: $d_1=166$ nm, $d_2=100$ nm, $d_3=117$ nm, $d_4=165$ nm, $d_5=174$ nm, $d_6=130$ nm, $d_7=168$ nm, and $d_8=176$ nm. In addition, the minimum distances between two consecutive sources has been set to  $1$ nm.

\section{Simulation Results and Discussion}
\label{sec:Simulation results and discussion}
This section discusses OOMMF simulation results for the Byte-based Majority gate. Subsequently, we compare the data parallel design with conventional SW gate based implementation and, finally, discuss scalability related issues and the effect of waveguide width scaling. 

\subsection{Simulation Results for $3$-input Majority Gate}

\begin{figure}[t]
\centering
  \includegraphics[width=0.5\linewidth]{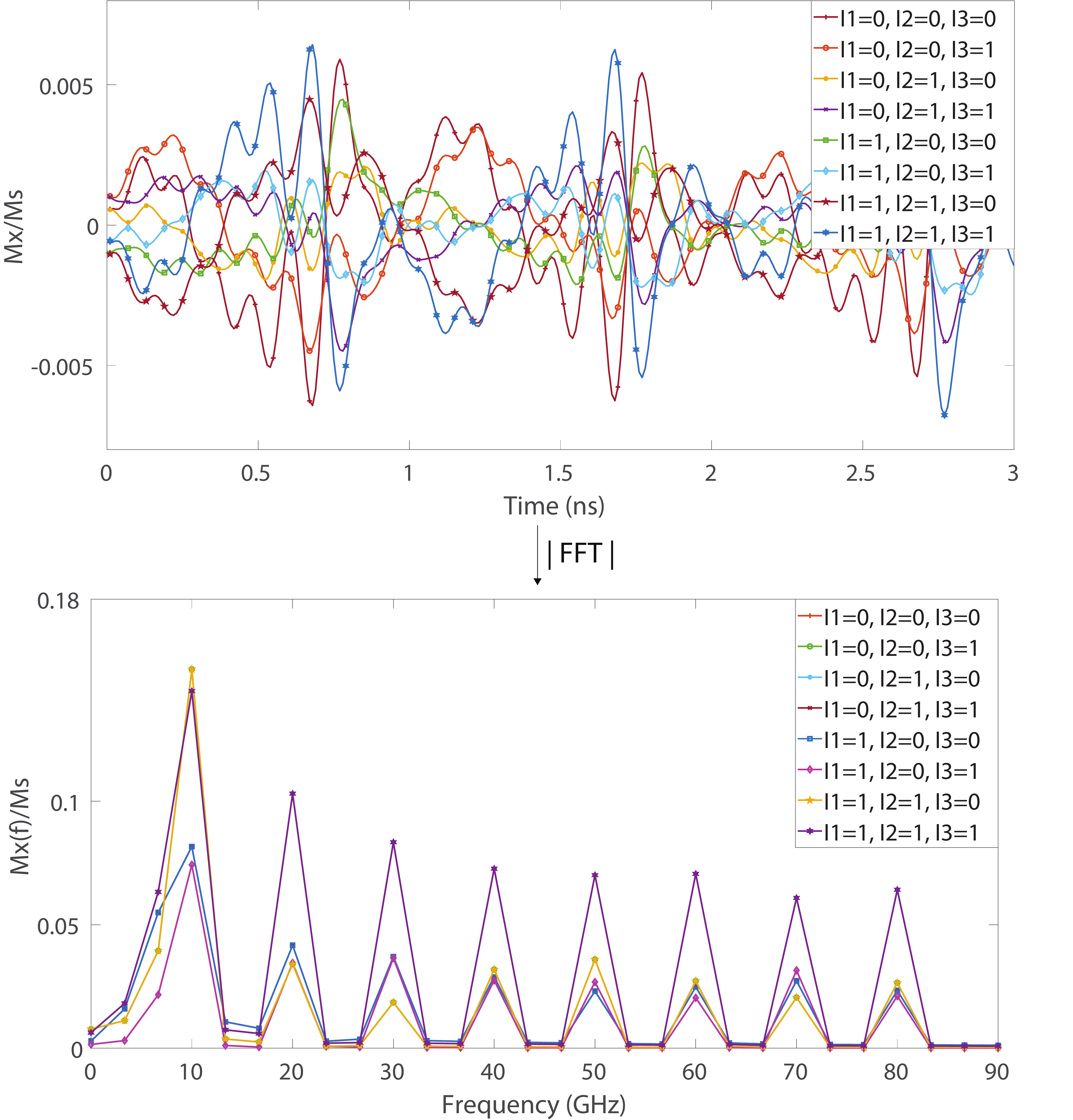}
  \caption{ Byte-Based Majority Gate Response in Time and Frequency}
  \label{fig:results3}
\end{figure} 

\begin{figure}[b]
\centering
  \includegraphics[width=0.5\linewidth]{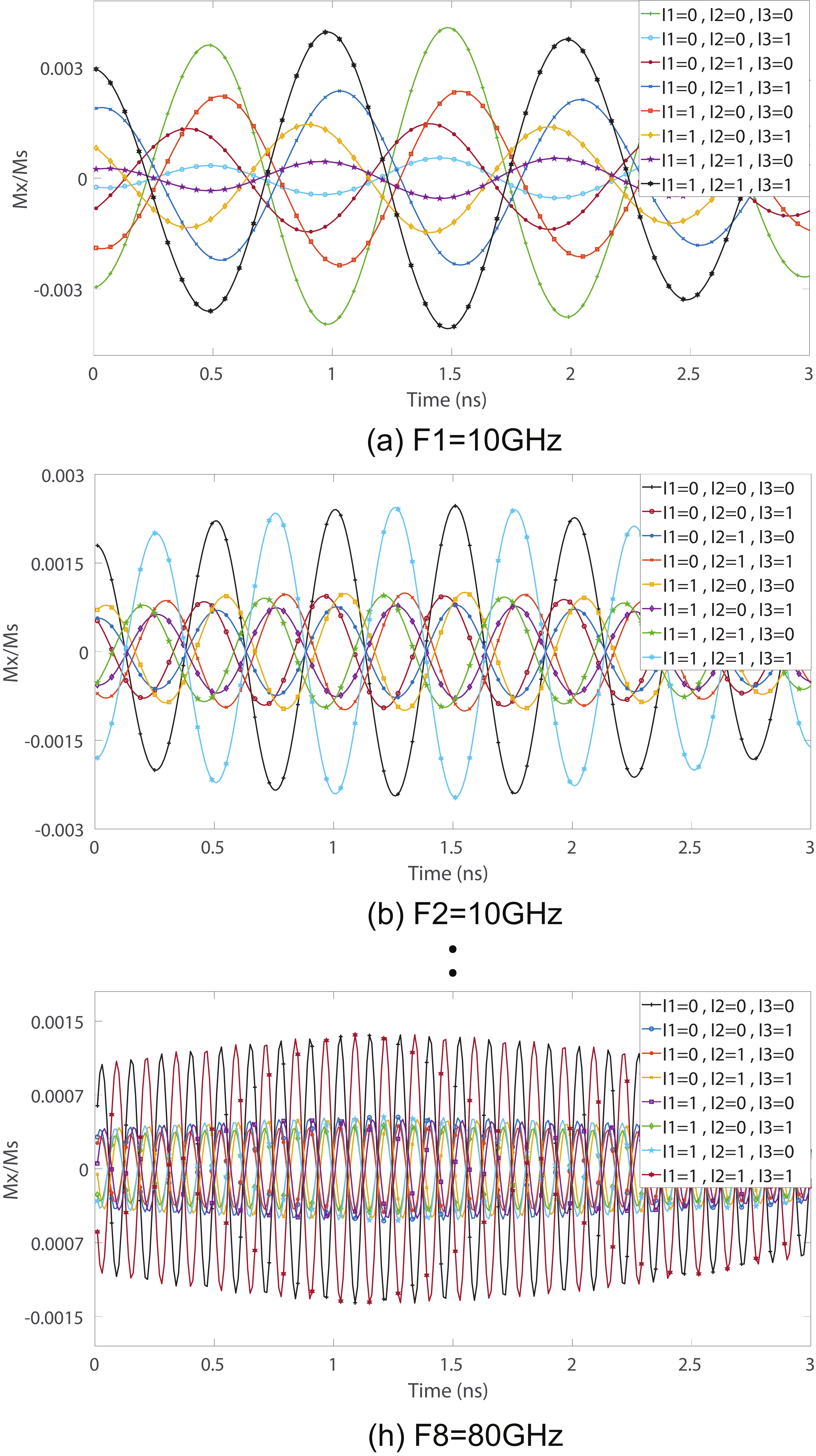}
  \caption{MAJ Gate Outputs: a) $f_1$=$10$GHz, b) $f_2$=$20$GHz, \ldots, h) $f_8$=$80$GHz}
  \label{fig:results4}
\end{figure} 

Figure \ref{fig:results3} presents OOMMF simulation results for the byte-based $3$-input Majority gate. Also in this case the $8$ output values can be extracted by taking the spin wave FFT amplitude absolute value and different frequency SWs do not affect each other as there are no frequency peaks at other frequencies than the excitation frequencies. As it can be observed in Figure \ref{fig:results4},  SWs with different frequencies propagate through the waveguide, and only SWs with similar frequency are interfering constructively and destructively depending on their phases. At the first output detector, which corresponds to a frequency $10$ GHz, the result of the $3$ inputs at this frequency is obtained as it is clear in Figure \ref{fig:results4} a). When $I_1= I_2=I_3=0$, the waves interfere constructively resulting in a phase $0$ SW, i.e., a logic $0$. When only of the inputs is $1$ while the others are $0$ the waves interfere constructively and destructively, which also results in a phase $0$ SW.  However, when two or more inputs are $1$ the gate output is a phase $\pi$ SW, i.e., a logic $1$. 
This holds true of all the $8$ output detectors embedded in the gates.

\subsection{Comparison and Discussion}

To get  inside on the potential implications of our proposal we compare byte-based $3$-input Majority gate with counterpart implementation obtained by the instantiation of $8$ normal (scalar) Majority gate, in terms of area, delay, and energy consumption. We assume that excitation/detection cell are $10$~nm $\times$ $50$~nm in all cases and that they are the main contributors to circuit delay and energy consumption. %{\bf We better give a reference for this.}  %The scalar implementation consist  we evaluated a byte-based 2-input XOR and 3-input Majority gate by implementing them using in-line structure, and duplicate the structure $8$ times to implement the $8$ bits. 
This means that the two implementation styles exhibits similar delay and energy consumption as they use the same number of sources and detectors. In terms of area, the conventional approach consumes $0.116$ $\mu m^2$  real estate to implement Majority gate as $8$ Majority gates are required to evaluate the $8$-bit $3$-inputs Majority gate. In contrast, the $8$-bit data parallel gate requires only $0.0279$ $\mu m^2$ to implement Majority gate as all the SW inputs are excited in the same waveguide. That indicates that the byte parallel approach requires $4.16$x less area for the Majority gate implementation. % implementation using the proposed structure compared to the conventional approach. %\medskip\\

\noindent\textbf{Scalability} While the proposed structure is generic and in principle functions correctly for a high number of inputs. However, due to the damping effect, SWs might have to be excited at different intensity levels if large number of inputs are required, such that the $I_n$ energy $<$ $I_{n-1}$ energy $<$ .. $<$ $I_1$ energy to guarantee the correct functionality of the structure. This variation in input energy can be used just if incorrect behaviour of the logic gate is noticed. Also, the source itself might be programmed to generate these different levels without a need for the input variation. %\medskip\\

\noindent\textbf{Waveguide Width Variation} To examine the waveguide with effect on gate functionality we scaled it up to $500$~nm. We noticed that the width scaling doesn't affect the functionality of the proposed structure, no crosstalk effects were present, and that the logic gates are still functioning correctly. In addition, as the width increases the ferromagnetic resonance frequency decreases, and by implication the first used frequency can assume a lower value.

\section{Conclusions}
\label{sec:Conclusion}

In this paper, we introduced a novel data parallel multi-frequency Spin Wave computation paradigm the associated generic in-line gate structure, which builds upon the selective interference of SWs with different  frequencies coexisting in the same waveguide.  To validate and evaluate the impact of our proposal, we implemented byte-wide $3$-input Majority gates operating on $8$ different frequency SWs, validated them by means of OOMMF simulations, and compared their area, delay, and energy consumption with conventional SW gate implementation counterparts.  Our evaluation indicated that the byte-wide approach provided a $4.16$x area reduction for the implementation of an 8-bit 3-input Majority gate without inducing any delay and power overhead.

\section*{Acknowledgement}
This work has received funding from the European Union's Horizon 2020 research and innovation program within the FET-OPEN project CHIRON under grant agreement No. 801055.


\begin{thebibliography}{0}%
\makeatletter
\providecommand \@ifxundefined [1]{%
 \@ifx{#1\undefined}
}%
\providecommand \@ifnum [1]{%
 \ifnum #1\expandafter \@firstoftwo
 \else \expandafter \@secondoftwo
 \fi
}%
\providecommand \@ifx [1]{%
 \ifx #1\expandafter \@firstoftwo
 \else \expandafter \@secondoftwo
 \fi
}%
\providecommand \natexlab [1]{#1}%
\providecommand \enquote  [1]{``#1''}%
\providecommand \bibnamefont  [1]{#1}%
\providecommand \bibfnamefont [1]{#1}%
\providecommand \citenamefont [1]{#1}%
\providecommand \href@noop [0]{\@secondoftwo}%
\providecommand \href [0]{\begingroup \@sanitize@url \@href}%
\providecommand \@href[1]{\@@startlink{#1}\@@href}%
\providecommand \@@href[1]{\endgroup#1\@@endlink}%
\providecommand \@sanitize@url [0]{\catcode `\\12\catcode `\$12\catcode
  `\&12\catcode `\#12\catcode `\^12\catcode `\_12\catcode `\%12\relax}%
\providecommand \@@startlink[1]{}%
\providecommand \@@endlink[0]{}%
\providecommand \url  [0]{\begingroup\@sanitize@url \@url }%
\providecommand \@url [1]{\endgroup\@href {#1}{\urlprefix }}%
\providecommand \urlprefix  [0]{URL }%
\providecommand \Eprint [0]{\href }%
\providecommand \doibase [0]{http://dx.doi.org/}%
\providecommand \selectlanguage [0]{\@gobble}%
\providecommand \bibinfo  [0]{\@secondoftwo}%
\providecommand \bibfield  [0]{\@secondoftwo}%
\providecommand \translation [1]{[#1]}%
\providecommand \BibitemOpen [0]{}%
\providecommand \bibitemStop [0]{}%
\providecommand \bibitemNoStop [0]{.\EOS\space}%
\providecommand \EOS [0]{\spacefactor3000\relax}%
\providecommand \BibitemShut  [1]{\csname bibitem#1\endcsname}%
\let\auto@bib@innerbib\@empty
%</preamble>
\end{thebibliography}%


\begin{thebibliography}{00}
%\bibitem{data1} N. D. Shah, E. W. Steyerberg, and D. M. Kent, “Big Data and Predictive Analytics Recalibrating Expectations,” JAMA, 2018.
\bibitem{data2} R. L. Villars, C. W. Olofson, and M. Eastwood, “Big data What it is and why you should care,” IDC, 2011.
%\bibitem{ITRS} S. Agarwal, G. Burr, A. Chen, S. Das, E. Debenedictis, M. P. Frank, P. Franzon, S. Holmes, M. Marinella, and T. Rakshit, “International Roadmap of Devices and Systems 2017 Edition: Beyond CMOS Chapter.” Sandia National Lab.(SNL-NM), Albuquerque, NM (United States), Tech. Rep., 2018.
\bibitem{cmosscaling2} D. Mamaluy and X. Gao, “The Fundamental Downscaling Limit of Field Effect Transistors,” Applied Physics Letters, vol. 106, no. 19, p. 193503, 2015.
\bibitem{cmosscaling3} B. Hoefflinger, Chips 2020: A Guide to the Future of Nanoelectronics. Springer Science and Business Media, 2012.
\bibitem{cmosscaling1} N. Z. Haron and S. Hamdioui, “Why is CMOS Scaling Coming to an End?” in Design and Test Workshop, 2008. IDT 2008. 3rd International. IEEE, 2008, pp. 98–103.
%\bibitem{survey1} K. Bernstein, R. K. Cavin, W. Porod, A. Seabaugh, and J. Welser, “Device and Architecture Outlook for Beyond CMOS Switches,” Proceedings of the IEEE, vol. 98, no. 12, pp. 2169–2184, Dec 2010.
\bibitem{survey2} D. E. Nikonov and I. A. Young, “Overview of Beyond-CMOS Devices and a Uniform Methodology for their Benchmarking,” Proceedings of the IEEE, vol. 101, no. 12, pp. 2498–2533, Dec 2013.
\bibitem{logic21} M. P. Kostylev, A. A. Serga, T. Schneider, B. Leven, and B. Hillebrands, “Spin-Wave Logical Gates,” Applied Physics Letters, vol. 87, no. 15, p. 153501, 2005. 
\bibitem{logic12} T. Schneider, A. A. Serga, B. Leven, B. Hillebrands, R. L. Stamps, and M. P. Kostylev, “Realization of Spin-Wave Logic Gates,” Applied Physics Letters, vol. 92, no. 2, p. 022505, 2008. 
\bibitem{logic11} K.S. Lee and S.K. Kim, “Conceptual Design of spin-wave Logic Gates Based on a Machzehnder-Type spin-wave Interferometer for Universal Logic Functions,” Journal of Applied Physics, vol. 104, no. 5, p. 053909, 2008. 
\bibitem{logic17} I. A. Ustinova, A. A. Nikitin, A. B. Ustinov, B. A. Kalinikos, and E. Lhderanta, “Logic Gates Based on Multiferroic Microwave Interferometers,” in 2017 11th International Workshop on the Electromagnetic Compatibility of Integrated Circuits (EMCCompo), July 2017, pp. 104– 107.
\bibitem{Magnonic_transistor} A. V. Chumak, A. A. Serga, and B. Hillebrands, "Magnon Transistor for All-Magnon Data Processing," Nature Communication, volume 5, pp. 4700, 2014.
\bibitem{logic24} B. Rana and Y. Otani, “Voltage-Controlled Reconfigurable Spin-Wave Nanochannels and Logic Devices,” Physical Review Applied, vol. 9, p. 014033, Jan 2018. 
\bibitem{logic1} A. Khitun and K. L. Wang, “Non-Volatile Magnonic Logic Circuits Engineering,” Journal of Applied Physics, vol. 110, no. 3, p. 034306, 2011.
\bibitem{logic19} K. Nanayakkara, A. Anferov, A. P. Jacob, S. J. Allen, and A. Kozhanov, “Cross Junction spin-wave Logic Architecture,” IEEE Transactions on Magnetics, vol. 50, no. 11, pp. 1–4, Nov 2014.
\bibitem{logic100} T. Fischer, M. Kewenig, D. A. Bozhko, A. A. Serga, I. I. Syvorotka, F. Ciubotaru, C. Adelmann, B. Hillebrands, and A. V. Chumak, "Experimental prototype of a spin-wave majority gate", Applied Physics Letter, Vol. 110, February 2017, pp.  152401-1-4.
\bibitem{logic101} F. Ciubotaru, G. Talmelli, T. Devolder, O. Zografos, M. Heyns, C. Adelmann, and I. P. Radu, "First experimental demonstration of a scalable linear majority gate based on spin waves", IEEE International Electron Devices Meeting (IEDM), January 2019, pp. 36.1.1-36.1.4.

\bibitem{parallel_data_processing1} A. Khitun, “Multi-frequency magnonic logic circuits for parallel data processing,” Journal of Applied Physics, vol. 111, no. 5, pp. 054307, 2012. 

\bibitem{Magnonic_crystals_for_data_processing} A. V. Chumak, A. A. Serga, and B. Hillebrands, “Magnonic Crystals for Data Processing,” Journal of Physics D: Applied Physics, vol. 50, no. 24, p. 244001, 2017. 

\bibitem{LL_eq} L. Landau and E. Lifshitz., “On the Theory of the Dispersion of Magnetic Permeability in Ferromagnetic Bodies,” Physikalische Zeitschrift der Sowjetunion, pp. 101– 114, 1935.
%\bibitem{G_eq} T. L. Gilbert, “A Phenomenological Theory of Damping in Ferromagnetic Materials,” IEEE Transactions on Magnetics, vol. 40, no. 6, pp. 3443– 3449, Nov 2004.
%\bibitem{dispersionrelation} B. A. Kalinikos, and A. N. Slavin, "Theory of dipole-exchange spin-wave spectrum for ferromagnetic films with mixed exchange boundary conditions", Journal  Physics C: Solid State Physics, Vol. 19, 1986, pp. 7013-7033.
%\bibitem{Magnetostatics_ref3} A. A. Serga, A. V. Chumak, and B. Hillebrands, “YIG Magnonics,” Journal of Physics D: Applied Physics, Volume 43, No. 26, p. 264002, 2010.
%\bibitem{Magnonics} V. V. Kruglyak, S. O. Demokritov, and D. Grundler, “Magnonics,” Journal of Physics D: Applied Physics, vol. 43, no. 26, p. 264001, 2010. [Online]. Available: http://stacks.iop.org/0022-3727/43/i=26/a=264001
%\bibitem{counter} P. SHABADI, S. N. RAJAPANDIAN, S. KHASANVIS, and C. A. MORITZ, “Design of spin-wave Functions-Based Logic Circuits,” SPIN, vol. 02, no. 03, p. 1240006, 2012. [Online]. Available: https://doi.org/10.1142/S2010324712400061
%\bibitem{logic9} O. Zografos, L. Amar, P. Gaillardon, P. Raghavan, and G. D. Micheli, “Majority Logic Synthesis for spin-wave Technology,” 17th Euromicro Conference on Digital System Design, AugUST 2014, pp. 691– 694.
\bibitem{OOMMF} M. J. Donahue and D. G. Porter, “OOMMF User’s Guide, version 1.0,” Interagency Report NISTIR 6376, Sept 1999.

\bibitem{parameters} T. Devolder, J.-V. Kim, F. Garcia-Sanchez, J. Swerts, W. Kim, S. Couet, G. Kar, and A. Furnemont, “Time-Resolved Spin-Torque Switching in MgO-Based Perpendicularly Magnetized Tunnel Junctions,” Physics Review B, vol. 93, p. 024420, Jan 2016.
%\bibitem{Excitation_table_ref16} O. Zografos, B. Sore, A. Vaysset, S. Cosemans, L. Amar, P. Gaillardon, G. D. Micheli, R. Lauwereins, S. Sayan, P. Raghavan, I. P. Radu, and A. Thean, “Design and Benchmarking of Hybrid CMOS-spin-wave Device Circuits Compared to 10nm CMOS,” in 2015 IEEE 15th International Conference on Nanotechnology (IEEE-NANO), July 2015, pp. 686–689.
\end{thebibliography}
\end{document}